\documentstyle[12pt,epsf]{article} 

\def\simg{{\ \lower-1.2pt\vbox{\hbox{\rlap{$>$}\lower6pt\vbox{\hbox{$\sim$}}}}\ }}
\def\siml{{\ \lower-1.2pt\vbox{\hbox{\rlap{$<$}\lower6pt\vbox{\hbox{$\sim$}}}}\ }} 
\def\bfnabla{\mbox{\boldmath $\nabla$}}

\def\bfsigma{\mbox{\boldmath $\sigma$}}
\def\als{\alpha_{s}}
\def\al{\alpha}
\def\lQ{\Lambda_{\rm QCD}}

\def\dsl{\,\raise.15ex\hbox{/}\mkern-13.5mu D}

\newcommand{\nn}{\nonumber}
\newcommand{\be}{\begin{equation}}
\newcommand{\ee}{\end{equation}}
\newcommand{\bea}{\begin{eqnarray}}
\newcommand{\eea}{\end{eqnarray}}

\newcommand{\Appendix}[1]%
    {%
     \section{#1}%
      }

\begin{document}\setlength{\unitlength}{1mm}

\begin{titlepage}
\begin{flushright}
\tt{TTP01-22}
\end{flushright}

\vspace{1cm}
\begin{center}
\begin{Large}
{\bf Renormalization group improvement of the NRQCD
Lagrangian and heavy quarkonium spectrum}\\[2cm]
\end{Large} 
{\large Antonio Pineda}\footnote{pineda@particle.uni-karlsruhe.de}\\
{\it Institut f\"ur Theoretische Teilchenphysik\\ 
        Universit\"at Karlsruhe, 
        D-76128 Karlsruhe, Germany}\\
\end{center}

\vspace{1cm}

\begin{abstract}
We complete the leading-log renormalization group scaling of the NRQCD
  Lagrangian at $O(1/m^2)$. The next-to-next-to-leading-log
  renormalization group scaling of the potential NRQCD Lagrangian (as
  far as the singlet is concerned) is also obtained in the situation
  $m\als \gg \lQ$. As a by-product, we obtain the heavy quarkonium
  spectrum with the same accuracy in the situation $m\als^2 \simg
  \lQ$. When $\lQ \ll m\als^2$, this is equivalent to obtain the whole
  set of $O(m\als^{(n+4)} \ln^n \als)$ terms in the heavy quarkonium
  spectrum. The implications of our results in the non-perturbative
  situation $m\als \sim \lQ$ are also mentioned.  \vspace{5mm} 
\\ 
PACS numbers: 12.38.Cy, 12.38.Bx, 12.39.Hg, 11.10.St
\end{abstract}

\end{titlepage}
\vfill
\setcounter{footnote}{0} 
\vspace{1cm}

\section{Introduction}

Heavy quark-antiquark systems near threshold are characterized by the
small relative velocity $v$ of the heavy quarks in their center of
mass frame. This small parameter produces a hierarchy of widely
separated scales: $m$ (hard), $mv$ (soft), $mv^2$ (ultrasoft), ... .
The factorization between them is efficiently achieved by using
effective field theories, where one can organize the calculation as
various perturbative expansions on the ratio of the different scales
effectively producing an expansion in $v$. The terms in these series
get multiplied by parametrically large logs: $\ln v$, which can also
be understood as the ratio of the different scales appearing in the
physical system. Again, effective field theories are very efficient in
the resummation of these large logs once a renormalization group (RG)
analysis of them has been performed. This will be the aim of this
paper for the cases of NRQCD \cite{NRQCD} and potential NRQCD (pNRQCD)
\cite{pNRQCD,long}\footnote{We will use here the pole mass as the expansion
parameter leaving aside in this paper any considerations about
renormalons.}.

First, we will obtain the RG improved matching coefficients of the
NRQCD Lagrangian at one loop and up to $O(1/m^2)$. Since, by
construction, the matching coefficients of HQET are equal to the
analogous ones of NRQCD, these can already be obtained from the
literature \cite{previous,BM}. Therefore, only the four-heavy fermion
matching coefficients need to be computed to obtain the complete
leading-log (LL) RG improvement of the NRQCD Lagrangian at
$O(1/m^2)$. We will perform such a calculation in this paper. For the
spin-dependent four-heavy fermion matching coefficients there already
exists a computation in Ref. \cite{CKO}. We differ with their
results. Our evaluation is relevant in the study of the situation
$m\als \sim \lQ$. On the one hand they could be used to improve the
accuracy of phenomenological studies or lattice simulations of NRQCD. On the other hand, this situation has been studied
within an effective field theory framework in \cite{m1} where the
matching between NRQCD and a Schr\"odinger like formulation has been
achieved in a controlled fashion. In particular, it is possible to
write the potentials as Wilson loops multiplied by the matching
coefficients inherited from NRQCD. The obvious application is that the
matching coefficients here computed are the ones that multiply the
Wilson loops in the non-perturbative potentials. This is especially
relevant now that the complete expression for the potential at
$O(1/m^2)$ in terms of Wilson loops is available \cite{m1,old,CKO}. In
particular, it would be welcome to have an updated evaluation of
the lattice analysis of the heavy quarkonium spectrum made in
Ref. \cite{lattice} taking into account the complete $O(1/m^2)$ potential as
well as the now known complete set of LL NRQCD matching coefficients.

In the situation when $m\als \gg \lQ$, the matching between NRQCD and
pNRQCD, i.e. the computation of the potentials, can be done
perturbatively. In this case ultrasoft gluons as well as the
quark-antiquark in an octet configuration do exist at the matching
scale between NRQCD and potential NRQCD producing further divergences.
By taking into account these divergences as well as the divergences
computed before we have obtained the next-to-next-to-leading-log
(NNLL) RG improved pNRQCD Lagrangian (as far as the singlet is concerned).

If we are in the situation $\lQ^3/(m\als)^2 \ll m\als^2$, the leading
solution of the spectrum corresponds to a Coulomb-type bound state and the
non-perturbative effects are corrections. In this situation, by using
the previous result of the NNLL RG improved pNRQCD Lagrangian, we are
able to obtain the heavy quarkonium spectrum with the same accuracy. 

If instead, we are in the situation $\lQ \ll m\als^2$, from our previous
  result, we are able to obtain the whole set of $O(m\als^{(n+4)} \ln^n
  \als)$ terms in the heavy quarkonium spectrum. There already exists an
  evaluation \cite{HMS} within the vNRQCD framework
  \cite{vNRQCD1,vNRQCD2,vNRQCD3} of the RG improved Heavy Quarkonium
  mass when $\lQ \ll m\als^2$. We agree for the spin-dependent terms
  (since we do for the spin-dependent potentials computed in Ref.
  \cite{vNRQCD2}) but differ for the spin-independent ones.

\section{NRQCD}
NRQCD has an ultraviolet (UV) cut-off $\nu_{NR}=\{\nu_p,\nu_s\}$
satisfying $mv \ll \nu_{NR} \ll m$. At this stage $\nu_p \sim
\nu_s$. $\nu_p$ is the UV cut-off of the relative three-momentum of
the heavy quark and antiquark. $\nu_s$ is the UV cut-off of the
three-momentum of the gluons and light quarks. This does not seem to
give problems at the order we are working at but one should be
eventually careful upon the possible gauge dependence of this
splitting.

Indeed, the above cutoffs plus the matter content of the theory given below correspond to our definition of NRQCD. Within the {\it threshold expansion} framework \cite{BS} this corresponds to integrate out the {\it hard} modes of QCD in order to obtain NRQCD. Unfortunately, NRQCD already contains non-physical degrees of freedom for the phase space region it is aimed to describe, which implies that the terms in the Lagrangian do not have a unique size nor, therefore, power counting (to avoid this problem is one of the motivations for the construction of pNRQCD, which will be done in the next section). Nevertheless, this poses no problem for the NRQCD running considered in this section.

The NRQCD Lagrangian including light fermions reads at $O(1/m^2)$ (up
to field redefinitions) \cite{NRQCD,previous,BM}
\be
\label{LagNRQCD}
{\cal L}={\cal L}_g+{\cal L}_l+{\cal L}_{\psi}+{\cal L}_{\chi}+{\cal
L}_{\psi\chi}
\,,
\ee
where $\psi$ is the Pauli spinor that annihilates the fermion, $\chi$
is the Pauli spinor that creates the anti-fermion, $i D_0=i\partial_0
-gA_0$, $i{\bf D}=i\bfnabla+g{\bf A}$, 
\be
\label{Lg}
{\cal L}_g=-\frac{1}{4}G^{\mu\nu a}G_{\mu \nu}^a +c_1^{g} {1\over 4
m^2}g f_{abc} G_{\mu\nu}^a G^{b\mu}{}_\alpha G^{c\nu\alpha},
\ee
\bea
\label{Ll}
{\cal L}_l&=&\sum_i \bar q_i i \dsl q_i + 
c_1^{ll}\displaystyle \frac{g^2}{8m^2}\sum_{i,j}\bar{q_i} T^a \gamma^\mu
 q_i \ \bar{q}_j T^a \gamma_\mu q_j  +
c_2^{ll}\displaystyle \frac{g^2}{8m^2}\sum_{i,j}\bar{q_i} T^a \gamma^\mu
 \gamma_5 q_i \ \bar{q}_j T^a \gamma_\mu \gamma_5 q_j 
\nn
\\ 
&&
+
c_3^{ll}\displaystyle \frac{g^2}{8m^2}\sum_{i,j}\bar{q_i}  \gamma^\mu
 q_i \ \bar{q}_j \gamma_\mu q_j +
c_4^{ll}\displaystyle \frac{g^2}{8m^2}\sum_{i,j}\bar{q_i} \gamma^\mu
\gamma_5
 q_i \ \bar{q}_j \gamma_\mu \gamma_5 q_j,
\eea
\bea
\label{Lhl}
{\cal L}_{\psi}&=&
\psi^{\dagger} \Biggl\{ i D_0
+ \, c_k{{\bf D}^2\over 2 m} + \, c_4{{\bf D}^4\over 8 m^3}
+ c_F\, g {{\bf \bfsigma \cdot B} \over 2 m}
\\ \nonumber
&& \qquad
+ c_D \, g { \left({\bf D \cdot E} - {\bf E \cdot D} \right) \over 8 m^2}
+ i c_S \, g { {\bf \bfsigma \cdot \left(D \times E -E \times D\right) }\over 8 m^2} \Biggr\} \psi
\\
\nn
&&
+c_1^{hl}\displaystyle\frac{g^2}{8m^2}\sum_i\psi^{\dagger}
  T^a \psi \ \bar{q}_i\gamma_0 T^a q_i +
c_2^{hl}\displaystyle\frac{g^2}{8m^2}\sum_i\psi^{\dagger}\gamma^\mu\gamma_5
T^a \psi 
  \ \bar{q}_i\gamma_\mu\gamma_5 T^a q_i 
\\
\nn
&&
+c_3^{hl}\displaystyle\frac{g^2}{8m^2}\sum_i\psi^{\dagger}
  \psi \ \bar{q}_i\gamma_0 q_i
+c_4^{hl}\displaystyle\frac{g^2}{8m^2}\sum_i\psi^{\dagger}\gamma^\mu\gamma_5
\psi 
  \ \bar{q}_i\gamma_\mu\gamma_5 q_i,
\eea
analogously for ${\cal L}_{\chi}$ and
\be
{\cal L}_{\psi\chi} =
  {d_{ss} \over m^2} \psi^{\dag} \psi \chi^{\dag} \chi
+ {d_{sv} \over m^2} \psi^{\dag} {\bfsigma} \psi
                         \chi^{\dag} {\bfsigma} \chi
+
  {d_{vs} \over m^2} \psi^{\dag} {\rm T}^a \psi
                         \chi^{\dag} {\rm T}^a \chi
+
  {d_{vv} \over m^2} \psi^{\dag} {\rm T}^a {\bfsigma} \psi
                         \chi^{\dag} {\rm T}^a {\bfsigma} \chi
\,.
\ee
We have also included the ${\bf D}^4/m^3$ term above since it will be necessary
in the evaluation of the heavy quarkonium mass once the power counting
is established. Moreover, we will consider that the kinetic term
matching coefficients are protected by reparameterization invariance
($c_k=c_4=1$) \cite{RI}, however, we will often keep them explicit for
tracking purposes. 

The NRQCD matching coefficients are functions of
$\nu_{NR}=\{\nu_p,\nu_s\}$. Somewhat by definition, the matching
coefficients of the bilinear in the heavy quark fields and of the pure
gluonic terms are just functions of $\nu_s$, i.e. $c=c(\nu_s,m)\equiv c(\nu_s)$. In any case, it will
explicitely come out from the calculation. The complete LL running of
these matching coefficients in the above basis (\ref{Lg}-\ref{Lhl}) have been calculated by Bauer and Manohar
\cite{BM} in the (background) Feynman gauge\footnote{We thank C. Bauer for
communication on this point.} (some partial previous results already
exist in the literature \cite{previous}). Therefore, in order to
complete the RG running of the NRQCD Lagrangian we only need to
compute the four-heavy-quark operators with LL accuracy. This will be
our aim in the following.

\medskip

A procedure to get the $\nu_{s}$ dependence of the NRQCD matching
coefficients is by using HQET-like rules in NRQCD (by this we mean to
perform the perturbative expansion in $1/m$ prior to the computation
of the Feynman integrals). In fact, in our case, at one loop, all the
dependence of the matching coefficients is only due to $\nu_s$ since
no $\nu_p$ dependence appears at one-loop, i.e. $d(\nu_p,\nu_s,m)\equiv
d(\nu_p,\nu_s) \simeq d(\nu_s)$. This will be discussed below within
pNRQCD.

Formally, we can write the NRQCD Lagrangian as an expansion in $1/m$ in the following way:
\be
{\cal L}_{\rm NRQCD} =\sum_{n=0}^{\infty}{1\over m^n}\lambda_n^BO_n^B
, 
\ee
where the above fields and parameters should be understood as bare and the renormalization group equations of the renormalized matching
coefficients read
\be
\nu_s {d \over d \nu_s}\lambda=B_{\lambda}(\lambda)
.
\ee
The RG equations have a triangular structure (the standard
structure one can see, for instance, in HQET RG evaluations, ie. for
the Lagrangian (\ref{LagNRQCD}) setting the heavy antiquark field to zero): 
\bea
\nu_s {d \over d \nu_s}\lambda_0&=&B_0(\lambda_0)
\,,
\nn
\\
\nu_s {d \over d \nu_s}\lambda_1&=&B_1(\lambda_0)\lambda_1
\,,
\nn
\\
\nu_s {d \over d \nu_s}\lambda_2&=&B_{2(2,1)}(\lambda_0)\lambda_2+B_{2(1,2)}(\lambda_0)\lambda_1^2
\,,
\eea
$$
\cdots\,,
$$
where the different B's can be power-expanded in $\lambda_0$
($\lambda_0$ corresponds to the marginal operators (renormalizable
interactions)). For NRQCD we have $\lambda_0=\al_s$ and
$\lambda_{1}=\{c_k, c_F\}$, $\lambda_{2}=\{c_1^g, c_D, c_S,
\{c^{ll}\},\{c^{hl}\}, \{d\} \}$.
 
At this stage, we would like to stress that we are working in a
non-minimal basis of operators for the NRQCD Lagrangian. Consequently,
the values of (some of) the matching coefficients are ambiguous (only some
combinations with physical meaning are unambiguous). In particular, some
of the matching coefficients could depend upon the gauge in which the
calculation has been performed. Therefore, it is important to perform
the matching calculation in the same gauge (at least for those operators
which could suffer the ambiguity). We will further discuss this point
latter on.

The RG equations
for the $\{c\}$ in the Feynman gauge can be read from Bauer and Manohar
results \cite{BM}. Because of latter comparison, we only explicitely write the
equation for $c_D$, it reads
\be
\nu_s {d\over d\nu_s}c_D={\alpha_{s}\over 4 \pi}\left[{4 C_A \over
    3}c_D-\left( {2 C_A \over 3}+{32 C_f \over 3}\right)c_k^2-{10 C_A
    \over 3}c_F^2+{8T_Fn_f \over 3}c_1^{hl}\right],
\ee
where, $T_F=1/2$, $C_f={N_c^2-1 \over 2N_c}$ and $C_A=N_c$. The explicit
expression for the $c_D$ RG equation depends on the gauge\footnote{The
renormalization group evolution of the one-heavy quark sector has also
been done in a minimal basis in Ref. \cite{BM} by elliminating the
operator multiplying $c_1^{hl}$. In that case the expression of $c_D$
obtained in Ref. \cite{BM} is indeed gauge independent.}.

The RG equations for the $d's$ in the Feynman
gauge are new and read\footnote{For the record, we also display the
non-equal mass case equations with the definitions $d/m^2 \rightarrow
d/(m_1m_2)$. The equations for $d_{ss}$ and $d_{sv}$ remain equal, for
$d_{vv}$ one has to change $c_F^2 \rightarrow c_F^{(1)}c_F^{(2)}$ and
the equation for $d_{vs}$ reads
\be
\nu_s {d\over d\nu_s}d_{vs}=\left(4C_f-{3 C_A \over
2}\right)\als^2c_k^2 
+ { 3 \over 4}\als^2C_A\left({m_1 \over m_2}c_D^{(2)}+{m_2 \over
m_1}c_D^{(1)}\right)
-{ 5 \over 4}c_k^2 \als^2C_A\left({m_1 \over m_2}+{m_2 \over
m_1}\right)
\,.
\ee
} 
\bea
\nu_s {d\over d\nu_s}d_{ss}&=&-2C_f\left(C_f-{C_A \over 2}\right)\als^2c_k^2
\,,
\nn\\
\nu_s {d\over d\nu_s}d_{sv}&=&0
\,,
\nn\\
\nu_s {d\over d\nu_s}d_{vs}&=&4\left(C_f-C_A\right)\als^2c_k^2 
+ { 3 \over 2}\als^2C_Ac_D
\,,
\nn\\
\nu_s {d\over d\nu_s}d_{vv}&=&-{C_A \over 2}\als^2c_F^{2}
\label{RGeqh}
\,.
\eea
These equations have been obtained by explicit computation in the
Feynman gauge by Signer \cite{Signer} within the threshold
formalism \cite{BS}. We have obtained them by using the results of
Ref. \cite{Match}, which were performed in the Feynman gauge, plus doing the
explicit calculation of the terms that depend on $c_D$ in NRQCD in the Feynman
gauge. This proves to be enough since the dependence on $c_k^2$ of
Eq. (\ref{RGeqh}) can be inferred from the results of Ref. \cite{Match}
once the dependence on $c_D$ is known (since the spin-dependent 
terms will depend on $c_F^2$). Both
calculations agree. Note that it was needed not to have $\nu_p$
dependence at one-loop in order the argument to go through.

\medskip

As we have mentioned we are not working in a minimal basis. This shows
up in the ambiguity of the value of the matching coefficients of some
operators. At the practical level, this means that they will depend on
the specific basis of operators we have taken for the NRQCD Lagrangian
and on the procedure used (in particular on the gauge). Therefore, if
working in a non-minimal basis, one should be careful and do the
matching using the same gauge for all the operators (or at least for
those that are potentially ambiguous).

For illustration,
let us consider the case without light fermions. In this case, $c_D$
and $d_{vs}$ are ambiguous but not an specific combination (see
Eq. (\ref{Dd2})). In particular, $c_D$ could be absorbed by other
matching coefficients by
using a field redefinition \cite{BM}. We can check these statements by doing the
calculation in the Coulomb gauge. In this case we obtain (no
non-trivial change is now required in the non-equal mass case for the RG
equations of the 
four-heavy fermion matching coefficients)
\bea
\nu_s {d\over d\nu_s}c_D({\rm Coulomb})&=&{\alpha_{s}\over 4 \pi}\left[{22 C_A \over
    3}c_D-\left( {32 C_A \over 3}+{32 C_f \over 3}\right)c_k^2-{10 C_A
    \over 3}c_F^2\right]
\,,
\nn\\
\nu_s {d\over d\nu_s}d_{vs}({\rm Coulomb})&=&\left(4C_f-{3 C_A \over
2}\right)\als^2c_k^2 
\label{RGeqhcoulomb}
\,.
\eea
One can see that, as far as the combination that appears in
Eq. (\ref{Dd2}) is concerned, the physical result is unchanged.

In the following we will use the Feynman gauge results for the NRQCD
matching coefficients.

\medskip

With the above results we have completed the RG equations of the NRQCD
Lagrangian at one loop at $O(1/m^2)$. In order to solve these
equations, we need the (tree-level) matching conditions of the matching
coefficients at some matching scale. We choose as the matching scale
$m$. The $\{c(m)\}$ can be read, for instance, from \cite{BM}. The tree-level matching conditions for
the four-heavy fermion operators read
\bea
d_{ss}(m)&=&-3C_f\left(C_f-{C_A \over 2}\right)\pi\als(m)
\,,
\nn\\
d_{sv}(m)&=&C_f\left(C_f-{C_A \over 2}\right)\pi\als(m)
\,,
\nn\\
d_{vs}(m)&=&3\left(C_f-{C_A \over 2}\right)\pi\als(m)
\,,
\nn\\
d_{vv}(m)&=&-\left(C_f-{C_A \over 2}\right)\pi\als(m)
\label{matcond}
\,.
\eea

We can then obtain the solution of the RG equations. We only
explicitely display those which are new or will be necessary later on (we
define $z=\left[{\als(\nu_s) \over \als(m)}\right]^{1 \over
\beta_0}\simeq 1 -1/(2\pi)\als(\nu_s)\ln ({\nu_s \over m})$,
$\beta_0={11 \over 3}C_A -{4 \over 3}T_Fn_f$)
\bea
c_F(\nu_s)&=&z^{-C_A}
\,,
\nn\\
c_S(\nu_s)&=&2z^{-C_A}-1
\,,
\nn\\
c_D(\nu_s)&=&
{9C_A \over 9C_A+8T_Fn_f}
\left\{
-\frac{5 C_A + 4 T_F n_f}{4 C_A + 4
T_F n_f} z^{-2 C_A} +
\frac{C_A +16 C_f - 8 T_F n_f}{2(C_A-2T_F n_f)}
\right.
\nn
\\
&&\qquad
+ \frac{-7 C_A^2 + 32 C_A C_f - 4 C_A T_F n_f +32 C_f T_F
n_f}{4(C_A + T_F n_f)(2 T_F n_f-C_A)} z^{4 T_F n_f/3 - 2C_A/3}
\nn
\\
&&
\qquad
\left.
+{8T_Fn_f \over 9C_A}
\left[
z^{-2C_A}+\left({20 \over 13}+{32 \over 13}{C_f \over
C_A}\right)\left[1-z^{-13C_A \over 6}\right]
\right]
\right\}
\,,
\nn\\
d_{ss}(\nu_s)&=&d_{ss}(m)+4C_f\left(C_f-{C_A \over 2}\right){\pi\over\beta_0}\als(m)\left[z^{\beta_0}-1\right]
\,,
\nn\\
d_{sv}(\nu_s)&=&d_{sv}(m)
\,,
\nn\\
d_{vs}(\nu_s)&=&d_{vs}(m)-\left(C_f-C_A\right){8\pi\over\beta_0}\als(m)
\left[z^{\beta_0}-1 \right]
\nn
\\
&&
-{27C_A^2 \over 9C_A+8T_Fn_f}{\pi \over \beta_0}\als(m)
\left\{
-\frac{5 C_A + 4 T_F n_f}{4 C_A + 4
T_F n_f}{\beta_0 \over \beta_0-2C_A}\left(z^{\beta_0-2 C_A}-1\right) 
\right.
\nn
\\
&&
\qquad
+
\frac{C_A +16 C_f - 8 T_F n_f}{2(C_A-2T_F n_f)}\left(z^{\beta_0}-1\right)
\nn
\\
&&
\qquad
+ \frac{-7 C_A^2 + 32 C_A C_f - 4 C_A T_F n_f +32 C_f T_F
n_f}{4(C_A + T_F n_f)(2 T_F n_f-C_A)}
\nn
\\
&&
\qquad
\qquad
\times
{3\beta_0 \over 3\beta_0+4T_Fn_f-2C_A}\left( z^{\beta_0+4 T_F n_f/3 - 2C_A/3}-1\right)
\nn
\\
&&
\qquad
+{8T_Fn_f \over 9C_A}
\left[{\beta_0 \over \beta_0-2C_A}\left(z^{\beta_0-2C_A}-1\right)
+\left({20 \over 13}+{32 \over 13}{C_f \over C_A}\right)
\right.
\nn
\\
&&
\qquad\qquad
\left.
\left.
\times
\left(
\left[z^{\beta_0}-1\right]-{6\beta_0 \over 6\beta_0-13C_A}
\left[z^{\beta_0-{13C_A \over 6}}-1\right]
\right)
\right]
\right\}
\,,
\nn\\
d_{vv}(\nu_s)&=&d_{vv}(m)+{C_A \over
\beta_0-2C_A}\pi\als(m)\left\{z^{\beta_0-2C_A}-1\right\}
\label{RGeqhs}
\,.
\eea
The $\{c\}$ matching coefficients can be found or deduced from the results
in Ref. \cite{BM}. The $\{d\}$ matching coefficients are new. For the
spin-dependent $\{d\}$ matching coefficients there already exists an evaluation \cite{CKO} but we
differ with their result.
This finishes the RG evaluation of the NRQCD Lagrangian at one-loop at
$O(1/m^2)$.

With the above results one can resum the large logs associated to the
hard scale by running down the factorization scale $\nu_s$ up to the
next relevant scale. 

\medskip

Finally note that it is very important to know the basis of
operators one has been working in NRQCD as well as in which gauge the
calculation has been performed. In practice this means that one should
make sure that $c_D$ and $d_{vs}$ have been computed in the same way in
order to obtain the correct result.

\section{pNRQCD}

The above results are also a necessary step towards the RG improvement
of pNRQCD when $m\als \gg \lQ$, which we consider in the following.
 
By integrating out some soft degrees in NRQCD one ends up in
pNRQCD. This latter theory is defined by the cut-off
$\nu_{pNR}=\{\nu_p,\nu_{us}\}$, where $\nu_p$ is
the cut-off of the relative three-momentum of the heavy quarks and is
such that $mv \ll \nu_p \ll m$ and $\nu_{us}$ is the cut-off of the
three-momentum of the gluons and light quarks with $mv^2 \ll \nu_{us} \ll mv$. In
principle, we do not rule out the option of correlating $\nu_p$ with
$\nu_{us}$ in order to efficiently perform the renormalization group
improvement at higher orders \cite{vNRQCD1}. Nevertheless, at the order we are working
with, we not need to specify any relation between $\nu_p$ and $\nu_{us}$
since the dependence on $\nu_p$ would be a subleading effect. Therefore,
in this paper, we will treat them as independent.

The pNRQCD Lagrangian reads as follows:
\begin{eqnarray}                         
& & {\cal L}_{\rm pNRQCD} =
{\rm Tr} \,\Biggl\{ {\rm S}^\dagger \left( i\partial_0 
- c_k{{\bf p}^2\over m} +c_4{{\bf p}^4\over 4m^3}
- V^{(0)}_s(r) - {V_s^{(1)} \over m}- {V_s^{(2)} \over m^2}+ \dots  \right) {\rm S}
\nonumber \\
&& \nonumber 
\qquad \qquad + {\rm O}^\dagger \left( iD_0 - c_k{{\bf p}^2\over m}
- V^{(0)}_o(r) 
+\dots  \right) {\rm O} \Biggr\}
\nonumber\\
& &\qquad + g V_A ( r) {\rm Tr} \left\{  {\rm O}^\dagger {\bf r} \cdot {\bf E} \,{\rm S}
+ {\rm S}^\dagger {\bf r} \cdot {\bf E} \,{\rm O} \right\} 
+ g {V_B (r) \over 2} {\rm Tr} \left\{  {\rm O}^\dagger {\bf r} \cdot {\bf E} \, {\rm O} 
+ {\rm O}^\dagger {\rm O} {\bf r} \cdot {\bf E}  \right\}  
\nonumber\\
& &\qquad- {1\over 4} G_{\mu \nu}^{a} G^{\mu \nu \, a},
\label{pnrqcdph}
\end{eqnarray}
where we have explicitly written only the terms relevant to the
analysis at the NNLL of the singlet sector; S and O are the singlet and octet field respectively.  
All the gauge fields in Eq. (\ref{pnrqcdph}) are functions of the  
center-of-mass coordinate and the time $t$ only.
For a more extensive discussion we refer the reader to Refs. \cite{long,logs}.

\subsection{Potentials}
We now display the structure of the matching potentials $V^{(0)}_s$, $V^{(0)}_o$, $V_s^{(1)}$ and $V_s^{(2)}$, 
which are the relevant ones to our analysis.  

{\it 1) Order $1/m^0$}. From dimensional analysis, $V^{(0)}_{s}(r)$ can only
have the following structure: 
\begin{equation}
V^{(0)}_{s} \equiv  - C_f {\alpha_{V_{s}} \over r},
\label{defpot0s}
\end{equation}
and similarly for the static octet potential:
\begin{equation}
V^{(0)}_{o} \equiv \left({C_A \over 2}-C_f \right){\alpha_{V_{o}} \over r}.
\label{defpot0o}
\end{equation}

{\it 2) Order $1/m$}. From dimensional analysis and time reversal, $V^{(1)}_s$ can only have the 
following structure: 
\be
{V^{(1)}_s \over m} \equiv -{C_fC_A D^{(1)}_s \over 2mr^2}.
\label{V1}
\ee

{\it 3) Order $1/m^2$}. At the accuracy we aim, $V^{(2)}_s$ has the structure  
\bea
{V^{(2)}_s \over m^2} &=& 
- { C_f D^{(2)}_{1,s} \over 2 m^2} \left\{ {1 \over r},{\bf p}^2 \right\}
+ { C_f D^{(2)}_{2,s} \over 2 m^2}{1 \over r^3}{\bf L}^2
+ {\pi C_f D^{(2)}_{d,s} \over m^2}\delta^{(3)}({\bf r})
\nn
\\
& & + {4\pi C_f D^{(2)}_{S^2,s} \over 3m^2}{\bf S}^2 \delta^{(3)}({\bf r})
+ { 3 C_f D^{(2)}_{LS,s} \over 2 m^2}{1 \over r^3}{\bf L} \cdot {\bf S}
+ { C_f D^{(2)}_{S_{12},s} \over 4 m^2}{1 \over r^3}S_{12}(\hat{\bf r}),
\label{V2}
\eea
where $S_{12}(\hat{\bf r}) \equiv 3 \hat{\bf r}\cdot\bfsigma_1 \, \hat{\bf r}\cdot\bfsigma_2 
- \bfsigma_1\cdot\bfsigma_2$ and ${\bf S} = \bfsigma_1/2 + \bfsigma_2/2$. 

The coefficients, $\alpha_{V_s}$, $D_s$, ... contain some $\ln r$
dependence once higher order corrections to their leading (non-vanishing) values are taken into account. 
In particular, we will have expressions like $\delta^{(3)}({\bf r})\ln^n r$. This is not 
a well-defined distribution and should be understood as the Fourier transform of $\ln^n 1/k$. 
Nevertheless, in order to use the same notation for all the matching
coefficients, and since it will be sufficient for the purposes of this
paper, resum the leading logs, we will use the expression $\delta^{(3)}({\bf r})\ln^n r$, 
although it should always be understood in the sense given above.

\subsection{RG equations}
Formally, we can write the pNRQCD Lagrangian as an expansion in
$1/r$($=1/r$, $p$) and
$1/m$ in the following way:
\be
{\cal L}_{\rm pNRQCD} =\sum_{n=-1}^{\infty}r^n{\tilde V}_n^{(B)}O_n^{(B)}
+{1 \over m}\sum_{n=-2}^{\infty}r^n{\tilde V}_n^{(B,1)}O_n^{(B,1)}
+{1 \over m^2}\sum_{n=-3}^{\infty}r^n{\tilde V}_n^{(B,2)}O_n^{(B,2)}
+O\left({1 \over m^3}\right)
, 
\ee
where the above fields and parameters should be understood as bare. As
for the renormalized quantities, we define $V$ as the potentials and ${\tilde V}$ as the (almost) dimensionless
constants in it. The latter are in charge of absorbing the divergences
of the effective field theory. Therefore, they will depend on $\nu_p$
and $\nu_{us}$. Note that the dependence on $\nu_s$ of the NRQCD
matching coefficients has to cancel in ${\tilde V}$ since the new
effective theory does not have any ultraviolet cutoff dependent on
$\nu_s$. This discussion completely fixes the procedure to obtain the
RG equations of the potentials: by studying the UV behavior of pNRQCD
it is possible to obtain the scale dependence of the potentials on
$\nu_p$ and $\nu_{us}$ and the independence on $\nu_s$ trivially sets the
$\nu_s$ scale (in-)dependence of the potentials. Being more specific,
the potentials have the following structure:
$$
{\tilde V}(d(\nu_p,\nu_s,m),c(\nu_s,m),\nu_s,\nu_{us},r)={\tilde
V}(\nu_p,m,\nu_{us},r) \equiv {\tilde V}(\nu_p,\nu_{us})
\,.
$$
This produces the following RG equations:
\bea
\label{nus}
&&\nu_s {d\over d\nu_s}{\tilde V}(d(\nu_p,\nu_s,m),c(\nu_s,m),\nu_s,\nu_{us},r)
\\
\nn
&&
\qquad\qquad
=
\left[\nu_s {\partial \over \partial \nu_s}+\nu_s \left({d\over d\nu_s}d\right){\partial \over
    \partial d}+\nu_s \left({d\over d\nu_s}c\right){\partial \over
    \partial c}\right]
{\tilde V}(d,c,\nu_s,\nu_{us},r)=0
\,,
\eea
\be
\label{nup}
\nu_p {d\over d\nu_p}{\tilde V}(d(\nu_p,\nu_s,m),\nu_s,\nu_{us},r)
=
\nu_p \left({d\over d\nu_p}d\right){\partial \over
    \partial d}
{\tilde V}(d,\nu_s,\nu_{us},r)
\,.
\ee
The first equation just reflects the independence of the potential on
$\nu_s$. At the practical level, with the accuracy we are working, it
is equivalent to set $\nu_s= 1/r$ up to factors of order one. 
The second equation tells us that the dependence on $\nu_p$ is
inherited from the (four-heavy fermion) NRQCD matching
coefficients.

One of our aims will be to obtain the heavy quarkonium spectrum with
NNLL accuracy when $\lQ^3/(m\als)^2 \ll m\als^2$. In that situation
the leading order solution corresponds to a Coulomb-type bound state
and, leaving aside non-perturbative corrections, a perturbative
expansion is licit. In order to achieve this goal we will need the RG
improvement of the pNRQCD Lagrangian for the singlet sector with the
same accuracy. Being more precise, we will need
\be
\nu {d\over d\nu}{\tilde V}^{(0)}_s \sim \als^4
\,,
\qquad
\nu {d\over d\nu}{\tilde V}^{(1)}_s \sim \als^3
\,,
\qquad
\nu {d\over d\nu}{\tilde V}^{(2)}_s \sim \als^2
\,,
\ee  
as well as (due to mixing)
\be
\nu {d\over d\nu}{\tilde V}^{(0)}_o \sim \als^2
\,,
\qquad
\nu {d\over d\nu}{\tilde V}_A \sim \als
\,,
\qquad
\nu {d\over d\nu}{\tilde V}_B \sim \als
\,,
\ee
within an strict expansion in $\als$. 

\medskip

Let us consider first the dependence on $\nu_p$. Eq. (\ref{nup}) tells
us that the dependence on $\nu_p$ appears due to the four-heavy
fermion matching coefficients. These first appear at
$O(1/m^2)$. Therefore, we only need to obtain the equation $\nu_p
{d\over d\nu_p}{\tilde V}^{(2)}_s \sim \als^2$. In fact, at least at
lowest non-vanishing order, only the delta potentials are 
dependent on $\nu_p$. Of this type is precisely the leading dependence on
$\nu_p$ of pNRQCD. It appears through (the iteration of) enough
singular potentials when performing standard quantum mechanics
perturbation theory. Explicit inspection shows that these kind of
effects first appear at $O(m\als^6)$ in a perturbative computation of
the mass (this argument is based on the singularity of the (iteration
of the) potentials
plus taking into account their leading non-vanishing power in the $\als$
expansion). Therefore, we obtain
\be
\label{nup1}
\nu_p {d\over d\nu_p}{\tilde V}^{(2)}_s =0+O(\als^3).  
\ee 
By using Eq. (\ref{nup}), this is equivalent to $\nu_p {d\over d\nu_p}d
=0+O(\als^3)$, which was needed previously in NRQCD. In principle, 
this result could also be proved by explicit inspection
on the possible diagrams at the quark-gluon level that could give
divergences proportional to $\nu_p$. The final conclusion is that we can neglect any dependence on $\nu_p$ at the order
we are working at in the potentials, i.e. ${\tilde V}(\nu_p,\nu_{us})
\simeq {\tilde V}(\nu_{us})$. Therefore, we only have to
compute the $\nu_{us}$ scale dependence.

\medskip

The $\nu_{us}$-scale dependence could be taken from the computation in
\cite{short,logs,RG,long} (see also \cite{KP1}) by keeping track of the dependence of the result on the
different ${\tilde V}$. Let us note that we only need to do a one-loop
computation in order to achieve the necessary accuracy (plus the
already known two-loop singlet static potential). This should be
compared with the one-, two- and three-loop calculations that seem to
be necessary if the calculation is performed at the quark-gluon level as
in Ref. \cite{vNRQCD2,vNRQCD3,HMS}.

Formally, the renormalization group equations of the renormalized matching
coefficients due to the $\nu_{us}$-dependence read 
\be
\nu_{us} {d \over d \nu_{us}}{\tilde V}=B_{{\tilde V}}({\tilde V})
.
\ee
From a practical point of view one can organize the RG equations
within an expansion in $1/m$.

At $O(1/m^0)$, the analysis corresponds to the study of the static
limit of pNRQCD, which has already been carried out in
Ref. \cite{RG}. We repeat the basic points here for ease of reference. Since ${\tilde V}_{-1}\not= 0$, there are
relevant operators (super-renormalizable terms) in the Lagrangian
and the US RG equations lose the triangular structure that we enjoyed for the
RG equations of $\nu_s$. Still,
if ${\tilde V}_{-1} \ll 1$, a perturbative calculation of the renormalization
group equations can be achieved as a double expansion in ${\tilde V}_{-1}$ and
${\tilde V}_0$ (for a similar discussion in the context of scalar $\lambda \phi^n$-like
theories 
see \cite{AtanceCortes}), where the latter corresponds to the marginal operators (renormalizable interactions)). 
At short distances ($1/r \gg \lQ$), the static limit of pNRQCD lives in this situation. Specifically, we
have ${\tilde V}_{-1}=\{\al_{V_s},\al_{V_o}\}$, that fulfills ${\tilde
V}_{-1} \sim \als(r) \ll 1$;
${\tilde V}_0=\als(\nu_{us})$  and ${\tilde V}_{1}=\{V_A,V_B\} \sim 1$. Therefore, 
we can calculate the anomalous dimensions order by
order in $\als(\nu_{us})$. In addition, we also have an expansion in
${\tilde V}_{-1}$. Moreover, the specific form of the pNRQCD
Lagrangian severely constrains the RG equations general structure. The
result obtained in Ref. \cite{RG} reads
\bea
\nu_{us} {d\over d\nu_{us}}\al_{V_{s}}
&=& 
{2 \over 3}{\alpha_{s}\over
  \pi}V_A^2\left( \left({C_A \over 2}
-C_f\right)\al_{V_o}+C_f\al_{V_s}\right)^3  
,
\nn \\
\label{RGeq}
\nu_{us} {d\over d\nu_{us}} \al_{V_{o}}&=& {2 \over 3} {\alpha_{s}\over \pi}
V_A^2\left(\left({C_A \over 2} -C_f\right)\al_{V_o}+C_f\al_{V_s}\right)^3  
,
\nn\\
\nu_{us} {d\over d\nu_{us}}\alpha_{s}&=&-\beta_0 {\als^2 \over 2\pi}
,
\nn\\
\nu_{us} {d\over d\nu_{us}} V_{A}&=& 0 
,
\nn
\\
\nu_{us} {d\over d\nu_{us}} V_{B}&=& 0
\,.
\eea

\medskip

At higher orders in $1/m$, we only need to consider the singlet
potentials. The same considerations than for the static limit apply
here as far as the non-triangularity of the RG equations is
concerned. At $O(1/m,1/m^2)$, we have the following matching coefficients:
${\tilde V}_{-2}^{(1)}=\{D_s^{(1)},c_k\}$ and ${\tilde
V}_{-3}^{(2)}=\{D_{1,s}^{(2)},D_{2,s}^{(2)},D_{d,s}^{(2)},D_{S^2,s}^{(2)},D_{LS,s}^{(2)},D_{S_{12},s}^{(2)}\}$, 
and we obtain 
\bea
\nu_{us} {d\over d\nu_{us}}C_AD_s^{(1)}
&=& 
{16 \over 3}{\alpha_{s}\over
  \pi}V_A^2c_k\left[ \left({C_A \over 2}
    -C_f\right)\al_{V_o}+C_f\al_{V_s}\right]
\left[
2C_f\al_{V_s}+\left({C_A \over
      2}-C_f\right)\al_{V_o}
\right]
,
\nn \\
\label{RGeqm}
\nu_{us} {d\over d\nu_{us}}D_{d,s}^{(2)}&=& 
{16 \over 3}{\alpha_{s}\over
  \pi}V_A^2c_k^2\left({C_A \over 2}
    -C_f\right)\al_{V_o}
,
\nn\\
\nu_{us} {d\over d\nu_{us}}D_{1,s}^{(2)}
&=& 
{8 \over 3}{\alpha_{s}\over
  \pi}V_A^2c_k^2\left[ \left({C_A \over 2}
    -C_f\right)\al_{V_o}+C_f\al_{V_s}\right]
\,,
\eea
and zero for the other matching coefficients (in particular for the
spin-dependent potentials). 

In a more formal way, Eq. (\ref{RGeqm}) has the following structure:
\bea
\nu_{us} {d\over d\nu_{us}}{\tilde V}_{-2}^{(1)}
&\sim& 
{\tilde V}_{0}{\tilde V}_{-2}^{(1)}{\tilde V}_{-1}^2{\tilde V}_{1}^2
,
\nn \\
\label{RGeqmf}
\nu_{us} {d\over d\nu_{us}}{\tilde V}_{-3}^{(2)}
&\sim& 
{\tilde V}_{0}{\tilde V}_{-2}^{(1)\,2}{\tilde V}_{-1}{\tilde V}_{1}^2
\,.
\eea
In fact, in general, we have the structure (${\tilde
V}_m^{(0)} \equiv {\tilde V}_m$) 
\be
\nu_{us} {d\over d\nu_{us}}{\tilde V}_{m}^{(n)}
\sim 
\sum_{\{n_i\}\{m_i\}}{\tilde V}_{m_1}^{(n_1)}{\tilde
V}_{m_2}^{(n_2)}\cdots{\tilde V}_{m_j}^{(n_j)}\,, \quad {\rm with}\quad  
\sum_{i=1}^jn_i=n\;,\, \sum_{i=1}^jm_i=m\,,
\ee
and one has to pick up the leading contributions from all the possible
terms. 

\medskip

Eqs. (\ref{nus}), (\ref{nup1}), (\ref{RGeq}) and (\ref{RGeqm}) provide
the complete set of RG equations at the desired order. By using
Eqs. (\ref{nus}) and (\ref{nup1}), we obtain
\be
{\tilde V}={\tilde V}(d(1/r),c(1/r),\nu_s=1/r,\nu_{us},r)
\,.
\ee
We now need the initial condition in order to solve the US RG equations,
i.e. the matching conditions. We fix the initial point at
$\nu_{us}=1/r$. In summary, we need to know the singlet static
potential with $O(\als^3)$ accuracy, the $1/m$ potential with
$O(\als^2)$ accuracy, the $1/m^2$ potentials and the singlet octet
potential with $O(\als)$ accuracy and $V_A$ with $O(1)$
accuracy at $\nu_{us}=1/r$. They read  
\bea
{\alpha}_{V_s}(r^{-1}) &=&\alpha_{\rm s}(r^{-1})
\left\{1+\left(a_1+ 2 {\gamma_E \beta_0}\right) {\alpha_{\rm s}(r^{-1}) \over 4\pi}\right.
\nonumber\\
&&
\left.
+\left[\gamma_E\left(4 a_1\beta_0+ 2{\beta_1}\right)+\left( {\pi^2 \over 3}+4 \gamma_E^2\right) 
{\beta_0^2}+a_2\right] {\alpha_{\rm s}^2(r^{-1}) \over 16\,\pi^2}
 \right\},
\label{newpot0i}\nn\\ 
D^{(1)}_s(r^{-1})&=&\alpha_{\rm s}^2(r^{-1}),
\label{Ds1i}\nn\\ 
D^{(2)}_{1,s}(r^{-1})&=&\alpha_{\rm s}(r^{-1}),
\label{Ds2i}\nn\\ 
D^{(2)}_{2,s}(r^{-1})&=&\alpha_{\rm s}(r^{-1}),
\label{Ds22i}\nn\\ 
D^{(2)}_{d,s}(r^{-1})&=& \alpha_{\rm s}(r^{-1})(2+c_D(r^{-1})-2c_F^2(r^{-1})) 
\nn
\\
&&
+{1 \over \pi}\left[ d_{vs}(r^{-1})+3d_{vv}(r^{-1}) 
+ {1 \over C_f}(d_{ss}(r^{-1})+3d_{sv}(r^{-1})) \right],
\label{Dd2i}\nn\\ 
D^{(2)}_{S^2,s}(r^{-1})&=&\alpha_{\rm s}(r^{-1})c_F^2(r^{-1}) - {3 \over 2\pi
C_f}(d_{sv}(r^{-1})+C_f d_{vv}(r^{-1})) ,
\label{Dss2i}\nn\\ 
D^{(2)}_{LS,s}(r^{-1})&=& {\alpha_{\rm s}(r^{-1}) \over 3}(c_S(r^{-1})+2c_F(r^{-1})), 
\label{DLs2i}\nn\\ 
D^{(2)}_{S_{12},s}(r^{-1})&=&\alpha_{\rm s}(r^{-1}) c_F^2(r^{-1}),
\label{Dsten2i}\nn\\
{\alpha}_{V_o}(r^{-1}) &=&\alpha_{\rm s}(r^{-1}),
\label{Voi}\nn\\
V_A(r^{-1})&=&1
\label{vai},
\eea
where $\beta_1=34/3C_A^2-4C_fT_Fn_f-20/3C_AT_Fn_f$ and the values of $a_1$ and $a_2$ have been computed in
Ref. \cite{FSP}. 
We now have all the necessary ingredients to solve the RG equations. 

Eqs. (\ref{RGeq}) and (\ref{RGeqm}) give rise to subleading effects
within an strict expansion in $\als$. Therefore, we can approximate
them to (if not displayed the RG equation remains equal)
\bea
\nu_{us} {d\over d\nu_{us}}\al_{V_{s}}
&=& 
{2 \over 3}{\alpha_{s}(\nu_{us})\over
  \pi} \left({C_A \over 2}\right)^3\alpha_{s}^3(r^{-1})
\,,
\nn \\
\nu_{us} {d\over d\nu_{us}}\al_{V_{o}}
&=& 0
\,,
\nn \\
\nu_{us} {d\over d\nu_{us}}C_AD_s^{(1)}
&=& 
{16 \over 3}{\alpha_{s}(\nu_{us})\over
  \pi}{C_A \over 2}
\left(C_f+{C_A \over
      2}\right)\als^2(r^{-1})
\,,
\nn \\
\nu_{us} {d\over d\nu_{us}}D_{1,s}^{(2)}
&=& 
{8 \over 3}{\alpha_{s}(\nu_{us})\over
  \pi}{C_A \over 2}
\als(r^{-1})
\,,
\nn\\
\label{RGeqmsimpl}
\nu_{us} {d\over d\nu_{us}}D_{d,s}^{(2)}&=& 
{16 \over 3}{\alpha_{s}(\nu_{us})\over
  \pi}\left({C_A \over 2}
    -C_f\right)
\als(r^{-1})
\,.
\eea

We can finally obtain the RG improved potentials for
the singlet: 
\bea
{\alpha}_{V_s}(\nu_{us}) &=&{\alpha}_{V_s}(r^{-1})+
{C_A^3\over
  6\beta_0}\als^3(r^{-1}) \log\left(
\alpha_{s}(r^{-1})\over \alpha_{s}(\nu_{us}) \right),
\label{newpot0}
\nn
\\ 
D^{(1)}_s(\nu_{us})&=&D^{(1)}_s(r^{-1})+
{16\over 3\beta_0}\left({C_A \over 2}+C_f \right)\alpha_{\rm s}^2(r^{-1})\log\left(
\alpha_{s}(r^{-1})\over \alpha_{s}(\nu_{us}) \right),
\label{Ds1}
\nn\\ 
D^{(2)}_{1,s}(\nu_{us})&=&D^{(2)}_{1,s}(r^{-1})+
{8C_A\over 3\beta_0}\alpha_{\rm s}(r^{-1})\log\left(
\alpha_{s}(r^{-1})\over \alpha_{s}(\nu_{us}) \right),
\label{Ds2}
\nn
\\ 
D^{(2)}_{2,s}(\nu_{us})&=&D^{(2)}_{2,s}(r^{-1}),
\label{Ds22}
\nn
\\ 
D^{(2)}_{d,s}(\nu_{us})&=& D^{(2)}_{d,s}(r^{-1})
+
{32\over 3\beta_0}\left({C_A \over 2}-C_f \right)\als(r^{-1})\log\left(
\alpha_{s}(r^{-1})\over \alpha_{s}(\nu_{us}) \right),
\label{Dd2}
\nn
\\ 
D^{(2)}_{S^2,s}(\nu_{us})&=&D^{(2)}_{S^2,s}(r^{-1}),
\label{Dss2}
\nn
\\ 
D^{(2)}_{LS,s}(\nu_{us})&=& D^{(2)}_{LS,s}(r^{-1}), 
\label{DLs2}
\nn
\\ 
D^{(2)}_{S_{12},s}(\nu_{us})&=&D^{(2)}_{S_{12},s}(r^{-1}).
\label{Dsten2}
\eea
This completes the RG evaluation of the pNRQCD Lagrangian at NNLL (as
far as the singlet is concerned).

\section{Heavy quarkonium spectrum}
In the situation $\lQ^3/(m\als)^2 \ll m\als^2$, the heavy quarkonium
behaves as a Coulomb-type bound state and the perturbative corrections can be
computed in a systematic way. From the potential--like terms,  
we obtain the following correction to the NNLO energy expression (the derivation of this
result would go along similar lines to those in Ref. \cite{logs}):
\bea
&& \delta E_{n,l,j}^{\rm pot}(\nu_{us}) = E_n \als^2  
\left\{-{2C_A \over 3\beta_0} \left[{C_A^2 \over 2} 
+4C_AC_f {1\over n(2l+1)}+2C_f^2\left({8 \over n(2l+1)} - {1 \over n^2}\right) \right]
 \log\left(
\alpha_{s}(\nu_{us})\over \alpha_{s} \right) \right.
\nn \\ 
\nn && 
\qquad
+{ C_f^2\delta_{l0} \over 3 n}
\left(
-{16\over \beta_0}\left[C_f-{C_A \over 2}\right]
\log\left(
\alpha_{s}(\nu_{us})\over \alpha_{s} \right) 
\right.
\nn
\\
&&\qquad\qquad
\left.
-{3\over 2}(1+c_D-2c_F^2)-{3 \over 2\pi\als}\left[ d_{vs}+3d_{vv} 
+ {1 \over C_f}(d_{ss}+3d_{sv}) \right]
\right)
\nn
\\ 
\label{energy1} 
&& \qquad  -{4 \over 3}{ C_f^2 \delta_{l0}\delta_{s1} \over n}   
\left\{
z^{-2C_A}-1 +{3 \over 2}{C_A \over \beta_0-2C_A}\left[z^{-\beta_0}-z^{-2C_A}\right]
\right\}
\\
\nn
&&
\left.\qquad
- 
{(1- \delta_{l0}) \delta_{s1} \over l(2l+1)(l+1)n}\,C_{j,l}{ C_f^2 \over 2} \right\} \,,
\eea
where $E_n= - mC_f^2\als^2/(4n^2)$, the scale $\nu_s$ in $z$ and in the
NRQCD matching coefficients has been fixed to
the soft scale $\nu_s=2a_n^{-1}$ where
$2a_n^{-1}={mC_f\als(2a_n^{-1}) \over n}$. $\als$ is also 
understood at the soft scale $\nu_s=2a_n^{-1}$ unless the scale is specified, and 
\begin{eqnarray} 
C_{j,l} = \,\left\{
\begin{array}{ll}
\displaystyle{ -{(l+1)\over 2\,l-1}\left\{4(2l-1)\left(z^{-C_A}-1\right)+\left(z^{-2C_A}-1\right)\right\}}&\ \ , \, j=l-1 \\
\displaystyle{-4\left(z^{-C_A}-1\right)+\left(z^{-2C_A}-1\right)}&\ \ ,\, j=l \\
\displaystyle{{l \over 2\,l+3}\left\{4(2l+3)\left(z^{-C_A}-1\right)-\left(z^{-2C_A}-1\right)\right\}}&\ \ ,\, j=l+1.
\end{array} \right.
\end{eqnarray}
Eq. (\ref{energy1}) gives all the $O(m\als^4(\als\ln)^n)$ terms for $n
\geq 1$ of the heavy quarkonium mass, where ln stands either for
$\ln(\als)$, arising from the hard scale, or for
$\ln(m\als/\nu_{us})$, arising from the ultrasoft scale. After adding
to Eq. (\ref{energy1}) the NNLO \cite{NNLO} result with the
normalization point at the {\it same} soft scale, $\nu_s=2a_n^{-1}$,
that we have used here, the complete (perturbative) NNLL heavy
quarkonium mass is obtained (note that for the LO result the
three-loop running $\als$ has to be used).

The $\nu_{us}$ dependence of Eq. (\ref{energy1}) 
cancels against contributions from US energies. At the next-to-leading order in the multipole
expansion, the contribution 
from these scales reads
\begin{equation}
\delta^{\rm US} E_{n,l}(\nu_{us}) = -i{g^2 \over 3 N_c}T_F  \int_0^\infty \!\! dt 
\langle n,l |{\bf r} e^{it(E_n-H_o)} {\bf r}| n,l \rangle \langle {\bf E}^a(t) 
\phi(t,0)^{\rm adj}_{ab} {\bf E}^b(0) \rangle(\nu_{us}), 
\label{energyNP}
\end{equation}
where $H_o = c_k\displaystyle{{\bf p}^2\over m} +V_o^{(0)}$ and $\nu_{us}$ is
the UV cut-off of pNRQCD. Then, the total correction to the energy reads  
\be
\delta E_{n,l,j}= \delta^{\rm pot} E_{n,l,j}(\nu_{us})+ \delta^{\rm US} E_{n,l}(\nu_{us})\,.
\label{energytotal}
\ee

Different possibilities appear depending on the relative size of $\lQ$ with
respect to the US scale $m\als^2$. If we consider that $\lQ \sim m\als^2$, the gluonic correlator in 
Eq. (\ref{energyNP}) cannot be computed using perturbation theory. Therefore,
in a model independent approach, one can leave it as a free parameter and fix 
it with experiment at some scale $\nu_{us}$ (since the running of
Eq. (\ref{energyNP}) with $\nu_{us}$ is known, one can then obtain its value at
another scale).

If we consider that $m\als^2 \gg \lQ$, Eq. (\ref{energyNP}) can be computed
perturbatively. Since $m\als^2$ is the next relevant scale, the effective role of
Eq. (\ref{energyNP}) will be to replace $\nu_{us}$ by $m\als^2$ (up to finite pieces that 
we are systematically neglecting) in Eq. (\ref{energy1}). Then
Eq. (\ref{energytotal}) reduces to Eq. (\ref{energy1}) with
$\nu_{us} \sim m\als^2$. In particular, we take $\nu_{us}=-E_n$. As expected, Eq. (\ref{energy1}) with
$\nu_{us}=-E_n$ reproduces the already known $O(m\als^5\ln\als)$
correction \cite{logs} (see also \cite{KP2,HMS}). Since in this situation one is assuming that $\displaystyle{\lQ \over m\als^2} \ll 1$, one
can expand on this parameter. Therefore, non-perturbative corrections can be
parameterized by local condensates. The leading and next-to-leading
non-perturbative corrections have been computed in the literature
\cite{VL,P1}. 

\medskip

There already exists an evaluation \cite{HMS} within the vNRQCD
framework \cite{vNRQCD1,vNRQCD2,vNRQCD3} of the RG improved Heavy Quarkonium
mass when $\lQ \ll m\als^2$. We agree for the spin-dependent terms
(since we agree with the spin-dependent potentials computed in
Ref. \cite{vNRQCD2}, see Eq. (\ref{Dsten2})) but
differ for the spin-independent ones. We note that the disagreement
still holds if we consider QED with light fermions ($C_f \rightarrow 1$,
$C_A \rightarrow 0$, $T_F \rightarrow 1$). Agreement is found for QED without light
fermions ($C_f \rightarrow 1$, $C_A \rightarrow 0$, $n_f \rightarrow
0$, $T_F \rightarrow 1$).

\medskip

{\bf Acknowledgments}\\
We thank M. Beneke and specially A. Signer for collaboration in the
early stages of this paper. We thank C. Bauer, A. Hoang and specially
J. Soto for discussions. We thank the IPPP at Durham for hospitality
while part of this work was carried out.

%%%%%%%%%%%%%%%%%%%%% BIBLIOGRAPHY %%%%%%%%%%%%%%%%%%%%%%%%%%%%%%%%%%%%


\begin{thebibliography}{99}

\bibitem{NRQCD} W.E. Caswell and G.P. Lepage, Phys. Lett. {\bf B167}, 437 (1986); 
 G.T. Bodwin, E. Braaten and G.P. Lepage, Phys. Rev. {\bf D51}, 1125
(1995); (E) ibid. {\bf D55}, 5853 (1997).

\bibitem{pNRQCD} A. Pineda and J. Soto, Nucl. Phys. {\bf B}
(Proc. Suppl.) {\bf 64}, 428 (1998); Phys. Rev. {\bf D59}, 016005 (1999).

\bibitem{long} N. Brambilla, A. Pineda, J. Soto and A. Vairo,
  Nucl. Phys. {\bf B566}, 275 (2000).

\bibitem{previous} E. Eichten and B. Hill, Phys. Lett. {\bf B243}, 427
(1990); A.F. Falk, B. Grinstein and M.E. Luke, Nucl. Phys. {\bf B357},
185 (1991); B. Blok, J.G. Korner, D. Pirjol and J.C. Rojas,
Nucl. Phys. {\bf B496}, 358 (1997). 

\bibitem{BM}  C. Bauer and A.V. Manohar, Phys. Rev. {\bf D57}, 337 (1998).

\bibitem{CKO} Y.-Q. Chen, Y.-P. Kuang and R.J. Oakes, Phys. Rev. {\bf
D52}, 264 (1995).

\bibitem{m1} N. Brambilla, A. Pineda, J. Soto and A. Vairo,
Phys. Rev. {\bf D63}, 014023 (2001); A. Pineda and A. Vairo,
Phys. Rev. {\bf D63}, 054007 (2001); (E) ibid. {\bf D64}, 039902
(2001).

\bibitem{old} E. Eichten and F.L. Feinberg, Phys. Rev. {\bf D23}, 2724
(1981); M.E. Peskin, in Proceeding of the 11th SLAC Institute, SLAC
Report No. 207, 151, edited by P. Mc Donough (1983); D. Gromes,
Z. Phys. {\bf C26}, 401 (1984); A. Barchielli, E. Montaldi and
G.M. Prosperi, Nucl. Phys. {\bf B296}, 625 (1988); (E) ibid. {\bf
B303}, 752 (1988); A. Barchielli, N. Brambilla and G. Prosperi, Nuovo
Cimento {\bf A103}, 59 (1990); A.P. Szczepaniak and E.S. Swanson, Phys. Rev. {\bf D55}, 3987 (1997).

\bibitem{lattice} G.S. Bali and P. Boyle, Phys. Rev. {\bf D59}, 114504
(1999).  

\bibitem{HMS}  A.H. Hoang, A.V. Manohar and I.W. Stewart,
Phys. Rev. {\bf D64}, 014033 (2001). 

\bibitem{vNRQCD1} M.E. Luke, A.V. Manohar, I.Z. Rothstein,
Phys. Rev. {\bf D61}, 074025 (2000).  

\bibitem{vNRQCD2} A.V. Manohar and I.W. Stewart, Phys. Rev. {\bf D62},
014033 (2000).  

\bibitem{vNRQCD3} A.V. Manohar and I.W. Stewart, Phys. Rev. {\bf D62},
074015 (2000); Phys. Rev. {\bf D63}, 054004 (2001).

\bibitem{BS} M. Beneke and V.A. Smirnov, Nucl. Phys. {\bf B522}, 321
(1998).  

\bibitem{RI} M.E.~Luke and A.V.~Manohar, Phys.\ Lett.\ {\bf B286}, 348
(1992).

\bibitem{Signer} A. Signer, private communication.

\bibitem{Match} A. Pineda and J. Soto, Phys. Rev. {\bf D58}, 114011 (1998).

\bibitem{logs}  N. Brambilla, A. Pineda, J. Soto and A. Vairo,
     Phys. Lett. {\bf B470} (1999) 215.

\bibitem{short} N. Brambilla, A. Pineda, J. Soto and A. Vairo,
     Phys. Rev. {\bf D60} (1999) 091502. 

\bibitem{RG} A. Pineda and J. Soto, Phys. Lett. {\bf B495}, 323 (2000).

\bibitem{KP1} B.A. Kniehl and A.A. Penin, Nucl. Phys. {\bf B563}, 200
  (1999).

\bibitem{AtanceCortes} M. Atance and J.L. Cortes, Phys. Rev. {\bf D56} (1997)
  3611. 

\bibitem{FSP} W. Fischler, Nucl. Phys. {\bf B129}, 157 (1977);
Y. Schr\"oder, Phys. Lett. {\bf B447}, 321 (1999); M. Peter,
Phys. Rev. Lett. {\bf 78}, 602 (1997).

\bibitem{NNLO} A. Billoire, Phys. Lett. {\bf B92}, 343 (1980);
S. Titard and F.J. Yndur{\'a}in, Phys. Rev. {\bf D49}, 6007 (1994);
A. Pineda and F.J. Yndur{\'a}in, Phys. Rev. {\bf D58}, 094022 (1998).

\bibitem{KP2} B.A. Kniehl and A.A. Penin, Nucl. Phys. {\bf B577}, 197
(2000). 

\bibitem{VL} M.B. Voloshin, Nucl. Phys. {\bf B154}, 365 (1979); 
  H. Leutwyler, Phys. Lett. {\bf B98}, 447 (1981).

\bibitem{P1} A. Pineda, Nucl. Phys. {\bf B494}, 213 (1997).
\end{thebibliography}
\end{document}